\documentclass[12pt,a4paper]{article}
\usepackage{keyval,graphicx}     

\begin{document}

\title{Hadron Multiplicities in p+p and p+Pb Collisions\\ at the LHC\thanks
{Supported by National Natural Science Foundation of China
(11247322/A050306) and Natural Science Foundation of Hebei Province
(A2012210043)}} \small{\date{}}
\author{\small {WANG Hongmin$^{1}$\thanks{E-mail:whmw@sina.com.cn},
 HOU Zhaoyu$^{2}$, SUN Xianjing$^{3}$}\\
$^{1}$\small{Physics Department, Academy of Armored Forces
Engineering of PLA,
Beijing 100072, China}\\
$^{2}$\small{Physics Graduate School, Shijiazhuang Railway
Institute, 050043, China}\\
$^{3}$\small{Institute of High Energy Physics, Chinese Academy of
Sciences, Beijing 100049, China}}
 \maketitle
{\textbf{Abstract:} Experiments at the Large Hadron Collider (LHC)
have measured multiplicity distributions in p+p and p+Pb collisions
at a new domain of collision energy. Based on considering an
energy-dependent broadening of the nucleon's density distribution,
charged hadron multiplicities are studied with the phenomenological
saturation model and the evolution equation dependent saturation
model. By assuming the saturation scale have a small dependence on
the 3-dimensional root mean square (rms) radius at different energy,
the theoretical results are in good agreement with the experimental
data from CMS and ALICE collaboration. Then, the predictive results
in p+p collisions at $\sqrt{s}=$ 14
TeV of the LHC are also given.}\\


{\textbf{1 Introduction}}
\\

Charged hadron multiplicities in relativistic heavy ion collisions
are of great significance as its variables are very sensitive to the
underlying mechanism involved in the nuclear collisions$^{[1]}$.
These can provide a unique opportunity to test the predictions of
quantum chromo-dynamics (QCD) and understand the partonic structure
of the colliding nuclei. Among the variables, the pseudo-rapidity
distribution and negative binominal distribution (NBD) are two
important and interesting variable quantities to be studied on both
experimental and theoretical sides in recent years$^{[2-7]}$. The
pseudo-rapidity distribution of charged hadron is one quantity
always used to investigate the particle production mechanism in high
energy hadron-hadron and nucleus-nucleus collisions.
 For pseudo-rapidity density is sensitive
to the initial conditions of system and the hadronic final state
interactions, the study of the charged hadron densities at
mid-rapidity can provide the relevant information on the interplay
between hard parton-parton scattering process and soft process. The
NBD is another quantity which played major role in describing
multiplicity distribution of produced charged particles. The NBD can
describe well the multiplicity distribution almost in all inelastic
high energy collision processes except for the data particularly at
the highest available collider energies. In this paper, the
pseudo-rapidity distribution and the NBD in p+p and p+Pb collisions
will be studied in the framework of the Color Glass Condensate.

At very high energies or small Bjorken-x, QCD predicts that high
density gluon in a hadron wave function form a new state, the Color
Glass Condensate (CGC). In this regime, the gluon density increases
inside the hadron wave functions and becomes very large in
comparison to all other parton species (the valence quarks), and the
sea quarks are suppressed by the coupling $\alpha_{s}$ since they
can be produced from the gluons by the splitting $g\rightarrow
q\bar{q}$.
 Now, CGC has become an effective theory in describing
the gluon content of a high energy hadron or nucleus in the
saturation regime and many saturation models have been
established$^{[8-12]}$. These saturation models can be divided into
two main kinds: the phenomenological saturation model and the
evolution equation dependent saturation model. The phenomenological
saturation models, such as the Golec-Biernat and
W$\mathrm{\ddot{u}}$sthoff (GBW) model$^{[8]},$ the Kharzeev, Levin
and Nardi (KLN) model$^{[9]}$ and the Kovchegov, Lu and Rezaeian
(KLR) model$^{[10]}$, are very simple and their dipole-proton
scattering amplitude can be given by an analytic expression. The
evolution equation dependent saturation models, such as the impact
parameter dependent saturation model (IP-Sat)$^{[11]}$ and the
running coupling Balitsky-Kovchegov (rcBK) model$^{[12]}$, are based
on the DGLAP or the  rcBK evolution equation and they are very
useful at small Bjorken-x or high collision energy. Having studied
charged particle multiplicity distribution with the GBW
model$^{[7]}$, we will investigate charged hadron pseudo-rapidity
distribution and the NBD with the phenomenological KLN model and the
evolution equation dependent rcBK model, respectively.

In order to give an accurate theoretical analysis, the nucleon's
density distribution in position space, which control the
unintegrated gluon distribution (UGD) through the saturation scale
$Q_{s}$, must be considered. It should be pointed out that, due to
gluon saturation, the width of the gluon distribution inside a
nucleon should grow with collision energy $\sqrt{s}$$^{[13]}$. This
will lead to a broadening of the nucleon's density distribution in
position space as $\sqrt{s}$ increases. Thus, we here consider an
energy-dependent broadening of the Gaussian nucleon thickness
function, and assume the saturation scale have a small dependence on
the 3-dimensional root mean square (rms) radius at different
collision energy$^{[14]}$. The outline of this paper is the
following. The theoretical method is given in
Sec. 2 and the results and discussion are  given in Sec. 3. \\

{\textbf{2 Method}}
\\

For hadron-hadron collisions, the multiplicity per unit rapidity and
per unit transverse area can be given by$^{[15]}$
$$\frac{dN}{d^{2}\mathbf{b}dy}=\frac{2\pi^{3}N_{c}}{N_{c}^{2}-1}\int d^{2}\mathbf{r}_{\bot}\int^{\infty}_{0}
\frac{d^{2}\mathbf{p}_{T}}{p^{2}_{T}}\int_{0}^{p_{T}}d^{2}\mathbf{k}_{T}\alpha_{s}(\mathrm{max}\{{\frac{(\mathbf{k}_{T}+\mathbf{p}_{T})^{2}}
{4},\frac{(\mathbf{k}_{T}-\mathbf{p}_{T})^{2}}{4}}\})$$
\begin{equation}
\times
\phi_{1}(x_{1},\frac{(\mathbf{k}_{T}+\mathbf{p}_{T})^{2}}{4},\mathbf{b})\phi_{2}(x_{2},\frac{(\mathbf{k}_{T}-\mathbf{p}_{T})^{2}}{4},
\mathbf{b}-\mathbf{r}_{\bot}),
\end{equation}
where $N_{c}=3$, $x_{1,2}=(p_{t}/\sqrt{s})\mathrm{exp}(\pm y)$,
$\mathbf{b}$ is the impact factor and $\mathbf{r}$ is the transverse
position of the gluon. The running coupling constant
$\alpha_{s}(k^{2})=\mathrm{min}\{\frac{4\pi}{\beta_{0}\mathrm{ln}[({k^{2}+\Lambda^{2}})/{\Lambda_{QCD}^{2}}]},0.5\}$
with $\beta_{0}=11-\frac{2}{3}n_{f}=9$ and
$\Lambda=\Lambda_{QCD}=0.2$ GeV. For the unintegrated gluon
distribution $\phi$, we will use the form given by the KLN
model$^{[9]}$ and the rcBK model$^{[12]}$.

In the KLN model, $\phi$ is taken to be
 \begin{equation}
 \phi(x,k^{2},\mathbf{b})=\frac{\kappa C_{F}Q_{s}^{2}}{2\pi^{3}\alpha_{s}(Q_{s}^{2})}\left
 \{\begin{array}{cc}
 \frac{1}{Q_{s}^{2}+\Lambda^{2}}, &  k\leq Q_{s}\\
 \frac{1}{k^{2}+\Lambda^{2}}, &  k>Q_{s}
\end{array},
\right.
 \end{equation}
where $C_{F}=({N_{c}^{2}-1})/{(2N_{c})}$ and $\kappa$ is a
normalization factor. The unintegrated gluon distribution depend on
transverse position through the saturation scale
\begin{equation}
Q_{\mathrm{s,p}}^{2}(x,\mathbf{b})=Q^{2}_{0}(\frac{T_{\mathrm{p}}(\mathbf{b})}{T_{\mathrm{p,0}}})(\frac{0.01}{x})^{\lambda},
\end{equation}
where $Q^{2}_{0}=2$ GeV$^{2}$, $\lambda=0.288$$^{[8]}$ and
$T_{\mathrm{p,0}}$ is taken as 1 fm$^{-2}$. For the nucleon
thickness function, the Gaussian form is used
\begin{equation}
T_{\mathrm{p}}(\mathbf{b})=\frac{e^{-b^{2}/(2B)}}{2\pi B}.
\end{equation}
 where the proton width parameter $B$ can be computed from$^{[13]}$
\begin{equation}
B(\sqrt{s})=\frac{\sigma_{\mathrm{in}}(\sqrt{s})}{14.30}\texttt{fm}^{2},
\end{equation}
and $\sigma_{\mathrm{in}}(\sqrt{s})$ is the inelastic scattering
cross section. In this paper, we assume the gluon saturation scale
have a small dependence on the 3-dimensional rms radius of the
proton$^{[14]}$
\begin{equation}
 Q^{2}_{\mathrm{s,p}}(\sqrt{s})= Q^{2}_{\mathrm{s,p}}(\sqrt{s_{0}})(\frac{\pi r^{2}_{\mathrm{rms},0}}{\pi
 r^{2}_{\mathrm{rms}}})^{1/\delta},
\end{equation}
where $\delta=0.8$$^{[14]}$ and the 3-dimensional rms radius
$r_{\mathrm{rms}}=\sqrt{<r^{2}>}=\sqrt{3B}$. In Table 1 we collect a
few representative values.
\begin{table}[htbp]
\centering
\begin{tabular}{l|c|c|c|c|c}
\hline $\sqrt{s}$/TeV &0.9 &2.36 &5.02 &7 &14\\ \hline
$\sigma_{\mathrm{in}}$/mb &52 &60 &67 &70.45 &76.3\\
$\sqrt{B}$/fm &0.603 &0.648 &0.685 &0.702 &0.730\\
$r_{\mathrm{rms}}$/fm &1.044 &1.192 &1.186 &1.216 &1.264\\ \hline
\end{tabular}
\caption{The 3-dimensional rms radius for various collision
energies. The values for $\sigma_{\mathrm{in}}$ at LHC energies were
reported in [6,16-17].}
\end{table}

In the rcBK model, the unintegrated gluon distribution can be
obtained from the dipole scattering amplitude via a Fourier
transform
\begin{equation}
\phi(x,k)=\int\frac{d^{2}\mathbf{r}}{2\pi r^{2}}e^{i\mathbf{k}\cdot
\mathbf{r}}N(x,r)=\int \frac{dr}{r}J_{0}(rk)N(x,r),
\end{equation}
where $J_{0}$ is the spherical bessel function of the first kind.
The dipole scattering amplitude in the rcBK evolution reads$^{[12]}$
\begin{equation}
\frac{\partial N(r,Y)}{\partial Y}=\int
d\mathbf{r}_{1}K^{\mathrm{Bal}}(\mathbf{r},\mathbf{r_{1}},\mathbf{r_{2}})
[N(r_{1},Y)+N(r_{2},Y)-N(r,Y)-N(r_{1},Y)N(r_{2},Y)],
\end{equation}
and the kernel for the running term using Balitsky's prescription
reads
\begin{equation}
K^{\mathrm{Bal}}(\mathbf{r},\mathbf{r_{1}},\mathbf{r_{2}})=\frac{N_{c}\alpha_{\mathrm{s}}(r^{2})}{2\pi^{2}}
[\frac{r^{2}}{r_{1}^{2}r_{2}^{2}}+\frac{1}{r_{1}^{2}}(\frac{\alpha_{\mathrm{s}}(r_{1}^{2})}{\alpha_{\mathrm{s}}(r^{2}_{2})}-1)
+\frac{1}{r_{2}^{2}}(\frac{\alpha_{\mathrm{s}}(r_{2}^{2})}{\alpha_{\mathrm{s}}(r^{1}_{2})}-1)].
\end{equation}
For the initial conditions, the GBW ansatz is used for the dipole
scattering amplitude$^{[8]}$
\begin{equation}
N^{\mathrm{GBW}}(r,Y=0)=1-\mathrm{exp}[-(\frac{r^{2}Q_{s0}^{2}}{4})^{\gamma}],
\end{equation}
where $\gamma=1$ and $Q_{s0}^{2}$ is the initial saturation scale
squared.

Now let us take into account the negative binomial distribution.
Negative binomial distribution is a general property of
multi-particle production process regardless of type of colliding
particles, and the negative binomial probability distribution for
obtaining $n$ charged particles in the final state is given as
follows:
\begin{equation}
P(n)=\frac{\Gamma(k+n)}{\Gamma(k)\Gamma(n+1)}(\frac{\bar{n}}{k})^{n}(1+\frac{\bar{n}}{k})^{-n-k},
\end{equation}
where the mean multiplicity $\bar{n}$ can be calculated in the
framework of CGC by integrating $y$ in Eq. (1). The quantity $k$,
which is the fluctuation parameter, can be estimated as a function
of the saturation scale$^{[6]}$
\begin{equation}
k=\kappa'\frac{N_{c}^{2}-1}{2\pi}Q_{\mathrm{s,p}}^{2}(y,\sqrt{s})\sigma_{\mathrm{in}}(\sqrt{s}),
\end{equation}
where $\kappa'$ is a normalization factor. Here, in order to get an
analytic expression between $Q_{\mathrm{s,p}}$ and $y$,
$Q_{\mathrm{s,p}}$ is reconsidered by substituting
$x_{1,2}=(Q_{\mathrm{s}}/\sqrt{s})e^{\pm y}$ into Eq. (3) as in Ref.
[8].\\

\textbf{3 Results and Discussion}
\\

To evaluate the pseudo-rapidity distribution, Eq. (1) should be
rewritten using the transformation
\begin{equation}
y(\eta)=\frac{1}{2}\mathrm{ln}\frac{\sqrt{\mathrm{cosh}^{2}\eta+m_{0}^{2}/p_{T}^{2}}+\mathrm{sinh}\eta}
{\sqrt{\mathrm{cosh}^{2}\eta+m_{0}^{2}/p_{T}^{2}}-\mathrm{sinh}\eta},
\end{equation}
and the Jacobian can be correspondingly written as
\begin{equation}
J(\eta)=\frac{\partial y}{\partial
\eta}=\frac{\mathrm{cosh}\eta}{\sqrt{\mathrm{cosh}^{2}\eta+m_{0}^{2}/p_{T}^{2}}}.
\end{equation}
where $m_{0}$ is the rest mass of particle, which corresponds to the
order of the scale $\Lambda_{QCD}$. With a $\chi^{2}$ analysis of
the experimental data$^{[18]}$, the factor $\kappa$ in Eq. (2) is
equal to 0.51 and 0.89 for the theory with and without considering
the rms radius dependent saturation scale, respectively. In Fig.1,
the theoretical results for pseudo-rapidity distribution of charged
hadrons in p+p collisions at $\sqrt{s}=$0.9 TeV(a),  2.36 TeV(b), 7
TeV(c) and 14 TeV(d) are shown. The solid and dashed curves are the
results of the KLN model with and without considering the rms radius
dependence of the saturation scale, respectively. The dotted curves
are the results of the rcBK model. The experimental data come from
CMS$^{[2-3]}$. It is shown that the theoretical results of the KLN
model considering the rms radius dependence are in good agreement
with the experimental data. For the effective value of Bjorken-x in
the rcBK model is $10^{-12}<x<0.01$, it is not valid at small
$\sqrt{s}$ or large pseudo rapidity. Thus, only the results at
$\sqrt{s}=7$ TeV fit well to the data with the rcBK model. The
results shown in Fig.1(d) are the predictive results for forthcoming
LHC experiment at $\sqrt{s}=14$ TeV, and it is shown that the
predictive results of the rcBK model are almost the same as those of
the KLN model.

Fig.2 shows the negative binominal distribution of charged hadron at
$|\eta|<0.5$. The figure captions in Fig.2 are the same as that in
Fig.1. The experimental data come from ALICE$^{[4]}$. The agreement
is seen to be quite well for the KLN model considered the rms radius
dependent saturation scale at $\sqrt{s}=$0.9 and 2.36 TeV and for
the rcBK model at $\sqrt{s}=$7 TeV. Here, it should be noted that
the agreement is seen to be not quite well for the theoretical
results of the KLN model at $\sqrt{s}=$7 TeV even after considering
the rms radius dependent saturation scale. Thus, we will give a
systematic analysis of all $\eta$ regions by considering the impact
factor dependence of the mean multiplicity and the quantity $k$ in
the near future.

Recently, the experimental data of the charged hadron multiplicity
in p+Pb collisions at $\sqrt{s}=5.02$ TeV are given by ALICE
collaboration$^{[5]}$. In order to get a further test of the theory,
the pseudo-rapidity distribution of charged hadrons in p+Pb
collisions are investigated. For the nuclear density distribution of
Pb, we use the Woods-Saxon distribution$^{[19]}$
\begin{equation}
\rho=\frac{\rho_{0}}{(1+\mathrm{exp}[(r-R)/a]},
\end{equation}
 where $\rho_{0}$ corresponds
to the nucleon density in the center of the nucleus, $R$ is the
nuclear radius and $a$ is the "skin depth". The theoretical results
are shown in Fig.3. The solid curve is the result of the KLN model
and the dashed curve is the result that we give in Ref. [7]. The
dotted and the dash-dotted curves are the results of DPMJET$^{[20]}$
and HIJING with gluon shadowing parameter $s_{g}=0.28$$^{[21]}$,
respectively. It is shown that the theoretical results of the KLN
model fit well to the new experimental data.

In summary, the pseudo-rapidity distribution and the NBD of charged
hadron in p+p and p+Pb collisions are studied with the
phenomenological KLN model and the rcBK model. By considering an
energy-dependent broadening of the nucleon's density distribution in
position space and the rms radius dependence of the gluon saturation
scale, it is found that the theoretical results are in good
agreement with the experimental data from CMS and ALICE. The
predictive results in p+p collisions at $\sqrt{s}=$14 TeV of the LHC
will be examined by the forthcoming experiment.

\begin{newpage}

\end{newpage}

\begin{newpage}

\begin{figure}
\centering
 \caption{Figure1: Pseudo-rapidity distribution of charged hadrons
in p+p collisions at $\sqrt{s}=0.9$ TeV (a), 2.36 TeV (b), 7 TeV (c)
and 14 TeV (d). The solid and dashed curves are the results of the
KLN model with and  without considering the rms radius dependent
saturation scale, respectively. The dotted curves are the results of
the rcBK model. The data are from CMS$^{[2-3]}$.}
\end{figure}

\begin{figure}
\centering \caption{Figure2: Negative binominal distribution in p+p
collisions at $\sqrt{s}=0.9$ TeV (a) , 2.36 TeV (b), 7 TeV (c) and
14 TeV (d). The figure captions are the same as that in Fig.1. The
data are form ALICE$^{[4]}$.}
\end{figure}

\begin{figure}
\centering \caption{Figure3: Pseudo-rapidity distribution of charged
hadrons in minimum bias p+Pb collisions at $\sqrt{s}=5.02$ TeV. The
curves are the result of the KLN model (solid curve), the result
that we give in Ref. [7] (dashed curve), the results given in Ref.
[20] (dotted curve) and Ref. [21] (dash-dotted curve). The data come
from ALICE$^{[5]}$.}
\end{figure}

\end{newpage}
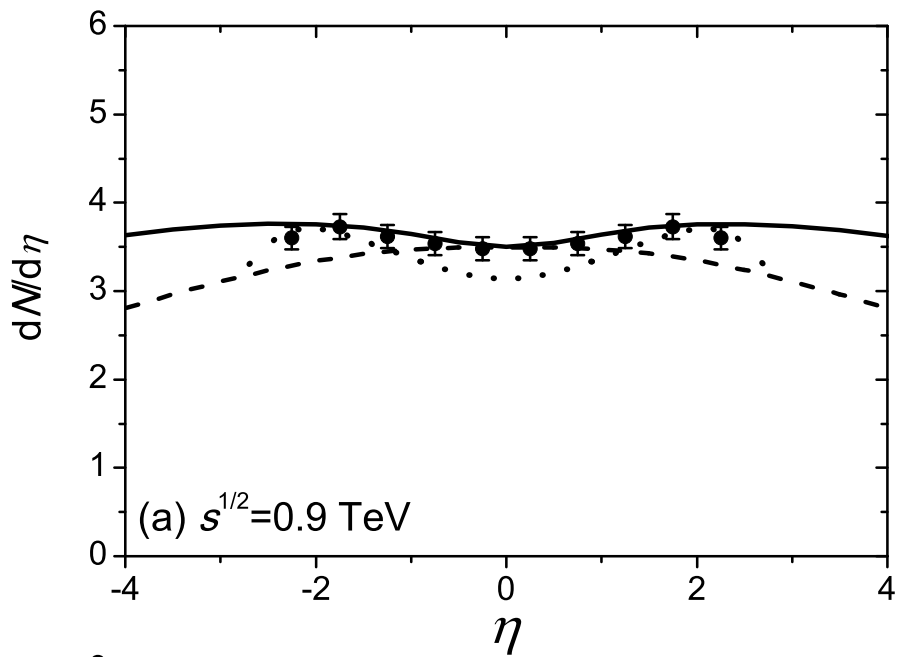
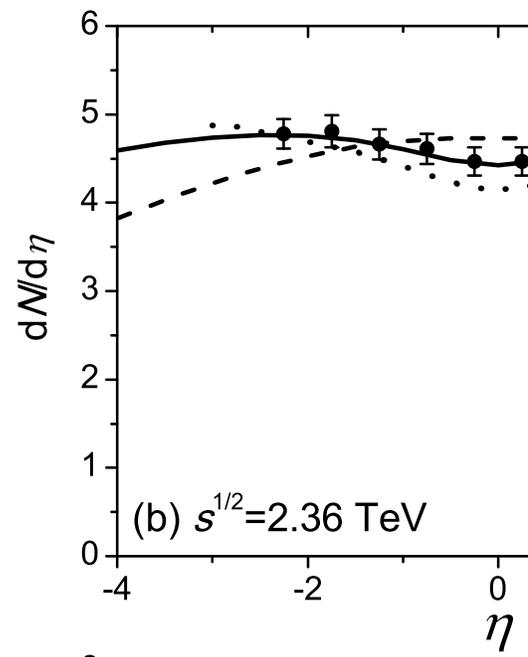
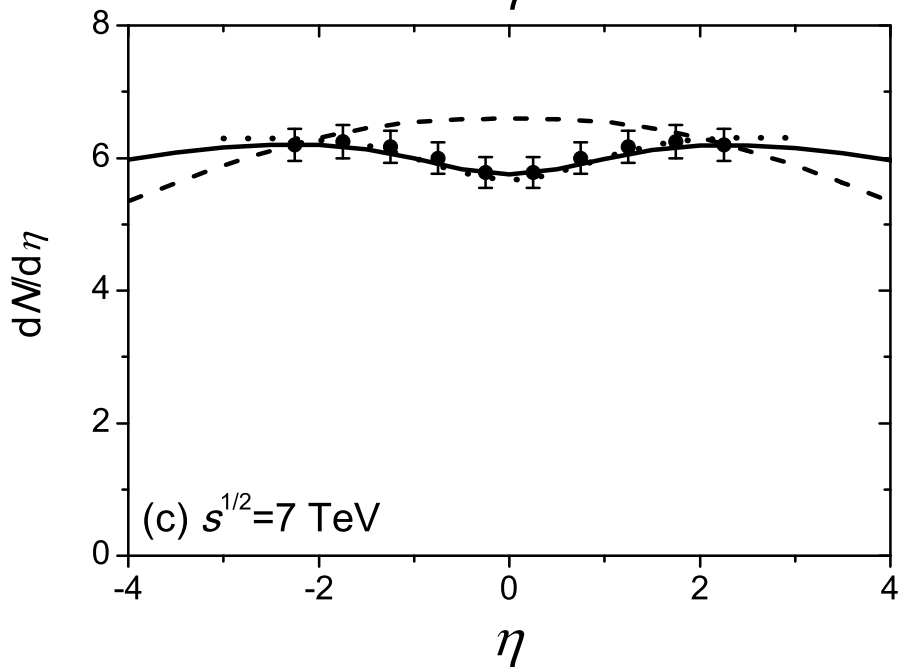
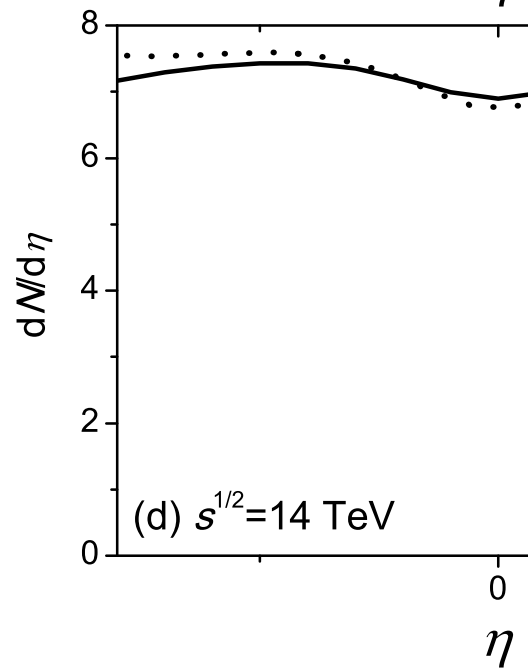
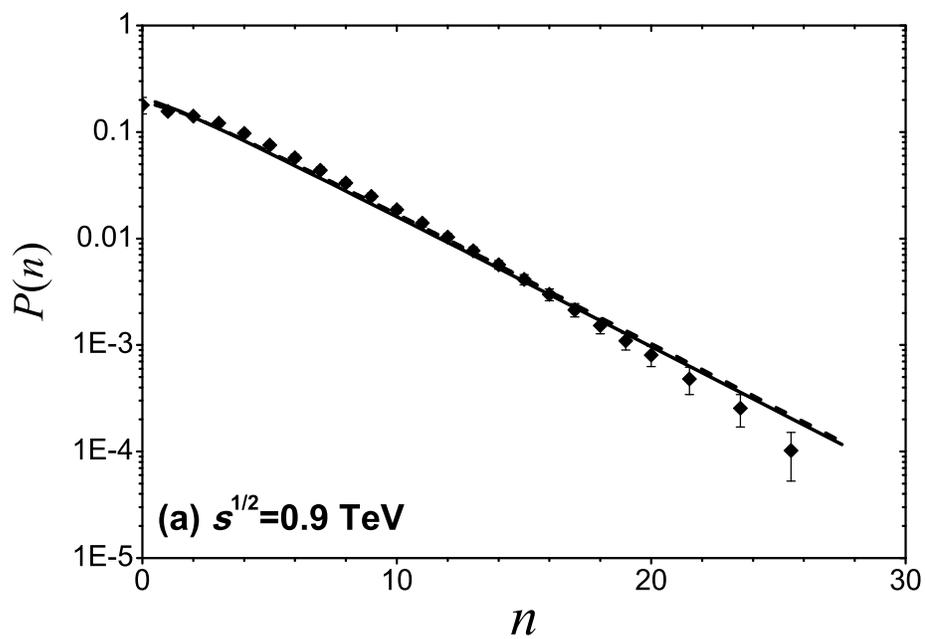
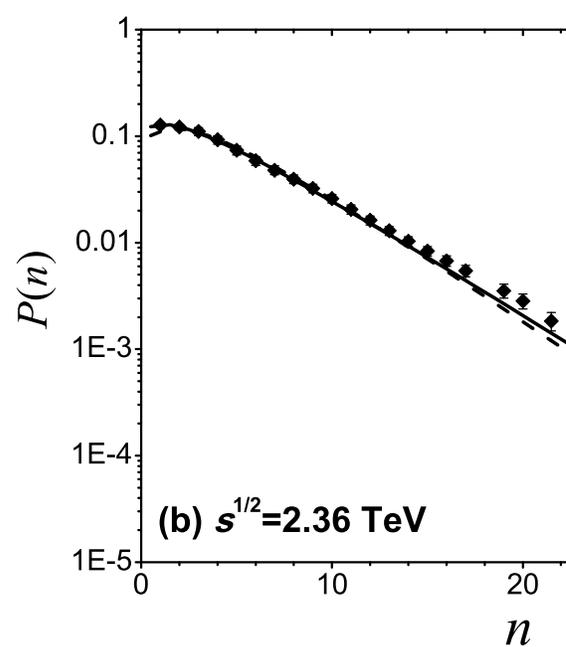
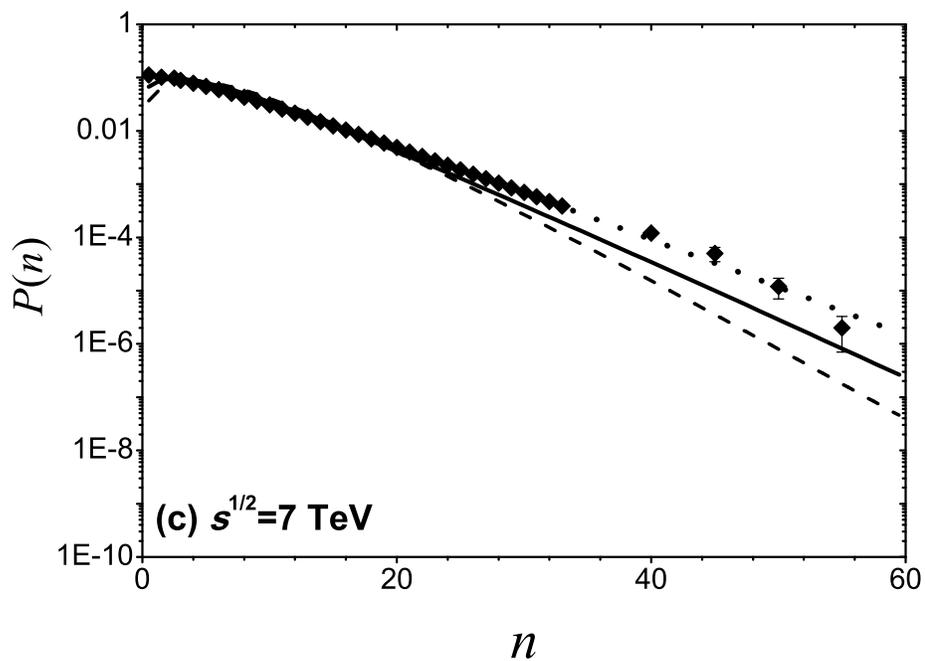
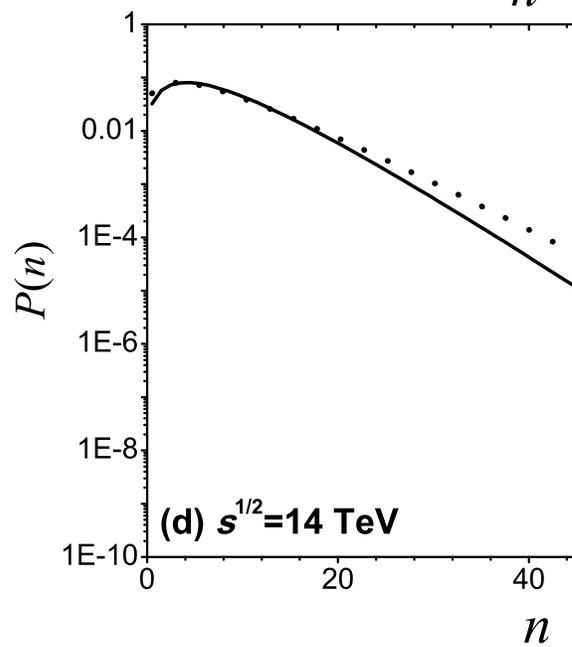
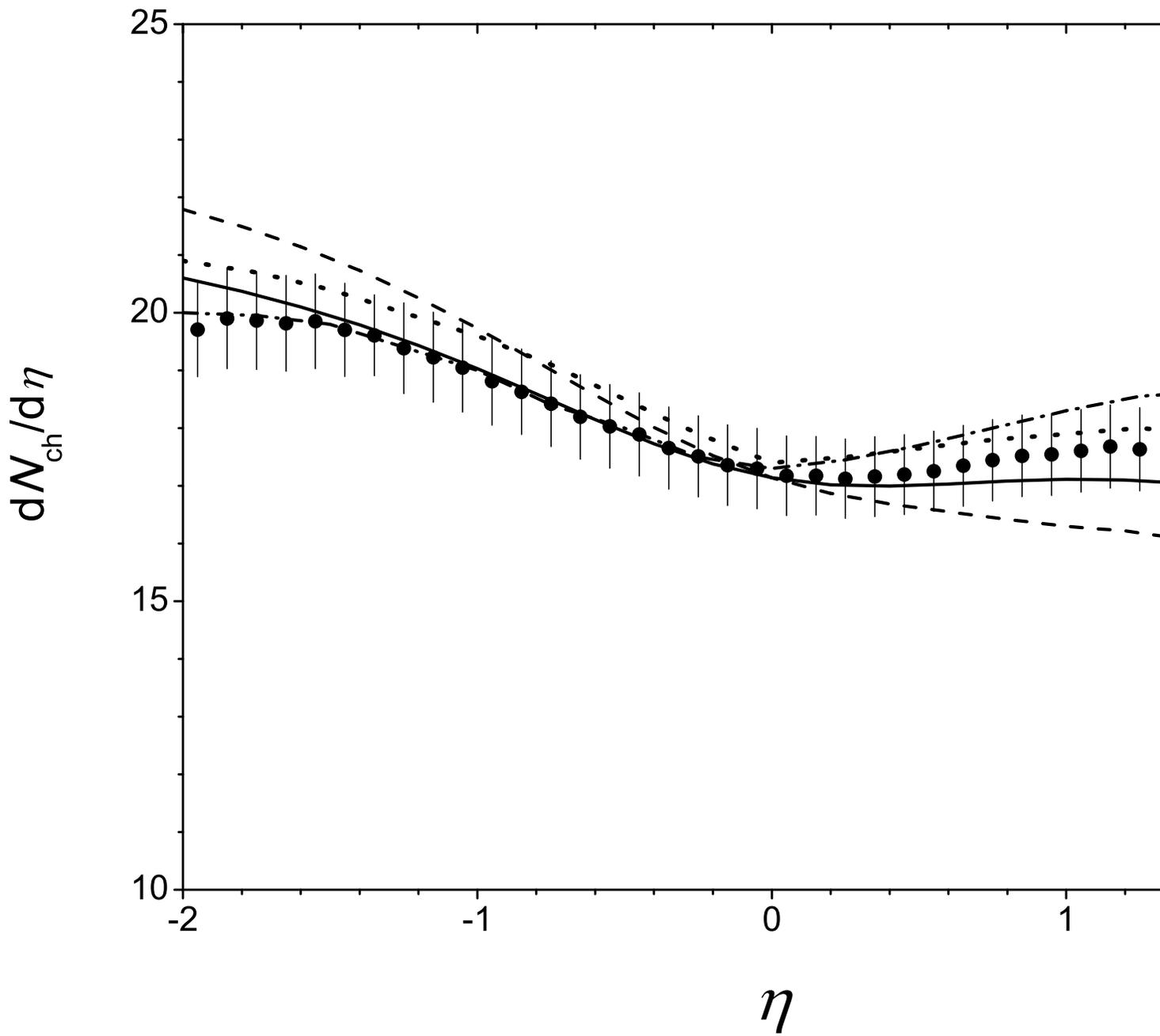
\end{document}